# Ultrafast Molecular Spectroscopy Using a Hollow-Core Photonic Crystal Fibre Light Source

Nikoleta Kotsina[1], Federico Belli[1], Shou-fei Gao[2], Ying-ying Wang[2], Pu Wang[2], John C. Travers[1] and Dave Townsend[1,3,a]

[1]*Institute of Photonics & Quantum Sciences, Heriot-Watt University, Edinburgh, EH14 4AS, UK*

[2]*Beijing Engineering Research Centre of Laser Technology, Institute of Laser Engineering, Beijing University of Technology, 100124, Beijing, China*

[3]*Institute of Chemical Sciences, Heriot-Watt University, Edinburgh, EH14 4AS, UK*

**Abstract**

We demonstrate, for the first time, the application of rare-gas filled hollow-core photonic crystal fibres (HC-PCFs) as tuneable ultraviolet light sources in femtosecond pump-probe spectroscopy. The time-resolved photoelectron imaging technique reveals non-adiabatic dynamical processes operating on three distinct timescales in the styrene molecule following 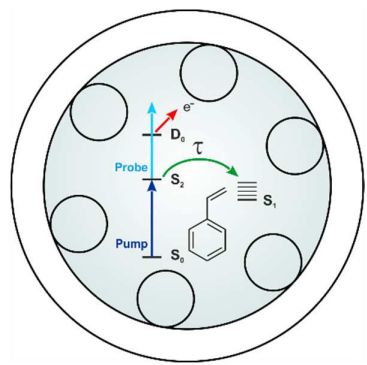 excitation over the 242-258 nm region. These include ultrafast (<100 fs) internal conversion between the $S_2(\pi\pi^*)$ and $S_1(\pi\pi^*)$ electronic states and subsequent intramolecular vibrational energy redistribution within $S_1(\pi\pi^*)$. Compact, cost-effective and highly efficient bench-top HC-PCF sources have huge potential to open up many exciting new avenues for ultrafast spectroscopy in the ultraviolet and vacuum ultraviolet spectral regions. We anticipate that our initial validation of this approach will generate important impetus in this area.

[a] *Corresponding author:* d.townsend@hw.ac.uk

Descriptions of dynamical behaviour in molecules are typically framed in terms of mechanistic pathways connecting sets of well-defined energy states. Underpinning this picture is the Born-Oppenheimer (BO) approximation, which assumes an adiabatic separation of nuclear motion (vibration/rotation) from electronic motion. This leads to the concept of electronic potential energy surfaces over which the much slower nuclear motion evolves. Breakdowns in the BO approximation occur when coupling between the nuclear and electronic degrees of freedom is not negligible, and such non-adiabatic processes are the rule, rather than the exception, in electronically excited molecules.[1-3] This is relevant to many fundamentally important processes including vision and photosynthesis, where BO breakdown effects are central to biological function.[4,5] They have also been implicated in so-called "self-protection" mechanisms taking place in, for example, DNA and the melanin pigmentation system, shielding living organisms from the potentially damaging effects of ultraviolet (UV) light.[6,7] It is therefore a hugely important challenge to develop a more detailed understanding of non-adiabatic processes and their relationship to chemical structure and mechanistic function.

Ultrafast femtosecond laser pulses with temporal durations comparable to the timescales of internal molecular motion are now widely employed in the study of non-adiabatic dynamics. The evolution of initially excited electronic states towards various photoproducts is followed in real time across multiple potential energy surfaces using pump-probe techniques. Time-resolved photoelectron imaging (TRPEI) is a powerful variant of this general approach, yielding highly differential energy- and angle-resolved information that offers deep insight into the complex molecular photophysics.[8-16] A key requirement for time-resolved electronic spectroscopy is the generation of broadly tuneable UV femtosecond pulses. Commercial femtosecond laser systems, however, typically produce output centred in the near infrared (NIR). Non-linear optical techniques are therefore required to access the UV region. Conventionally this is done using optical parametric amplifiers in conjunction with multi-stage frequency doubling or sum/difference frequency mixing in thin birefringent crystals. This is extremely inefficient (approx. 0.1-0.5%, relative to the infrared input) and, although extensively tuneable, is limited to ≥200 nm. High order non-linear processes such as four-wave mixing may be used in conjunction with gaseous media to obtain vacuum ultraviolet (VUV) pulses, although this is often even more inefficient (approx. 0.01-0.05%) and has limited ease of tuneability.[17-19]

An improved approach to producing UV/VUV femtosecond pulses for use in time-resolved spectroscopy is clearly desirable. A recently established route for achieving this goal is through optical soliton driven resonant dispersive wave (RDW) emission in gas-filled hollow fibres, where generation of tuneable femtosecond pulses covering the VUV to visible spectral region has been demonstrated.[20-26] This has been predominantly achieved using anti-resonant guiding hollow-core photonic crystal fibres (HC-PCFs), in which the required NIR pump source is just a few microjoules of energy for <50 fs pulses around 800 nm. In the time domain the process proceeds as follows: a single pump pulse is coupled into a gas-filled HC-PCF (see Fig. 1a). It experiences self-phase modulation (SPM) due to the intensity dependent refractive index of the filling gas, causing spectral broadening (Fig. 1b). If the gas pressure is adjusted such that the total (gas and waveguide) group velocity dispersion is anomalous at the input frequency, then the linear dispersion compensates the non-linear chirp due to SPM, leading to pulse compression. This subsequently enhances SPM and the cycle repeats leading to continuous (soliton) self-compression.[20, 21] This process can extend the pulse spectrum by over an octave, at which point higher-order dispersion can lead to a phase-matched (or resonant) transfer of energy from the self-compressing pulse to a specific frequency band (Fig. 1b). The precise dispersion tuneability of gas-filled HC-PCFs, obtained by adjusting the gas pressure, allows for a fine control of a phase-matched nonlinear process such as RDW emission.[20, 21] At reasonable gas pressures (1-30 bars) it has been experimentally tuned between 113 nm and 550 nm (Fig. 1c).[21, 22, 24, 25, 27] The emission is in the fundamental mode and can contain over 10% of the input 800 nm energy. As the process is transient, the emission is inherently ultrafast, with a duration as short as 3 fs,[28] and the low driving energy required makes it possible to scale the repetition rate of the UV emission to the MHz range.[29] Furthermore, the output polarization state closely follows that of the input pulse[26] – a particularly important consideration for the angle-resolved imaging aspect of TRPEI, where well-defined (linear) polarization is required.

Although the generation of femtosecond UV/VUV pulses using RDW emission in HC-PCFs is becoming well established, and has been demonstrated for use in angle-resolved photoemission spectroscopy of topological insulators,[30] the technique has not yet been exploited for use in time-resolved spectroscopy and molecular dynamics. Here we demonstrate the integration of an HC-PCF source into a TRPEI set-up for the first time, performing pump-probe measurements on styrene with 242-258 nm excitation. Styrene (see Fig. 2 for structure and UV absorption spectrum) has been studied previously using time-resolved photoelectron spectroscopy in this UV pump region and is known to exhibit ultrafast, spectrally well-resolved internal conversion between the $S_2(\pi\pi^*)$ and $S_1(\pi\pi^*)$ electronic states.[31, 32] It therefore provides an excellent system to demonstrate the first use of HC-PCF technology. Furthermore, there are discrepancies between these previous studies (expanded upon later) that our new measurements aim to resolve.

Our TRPEI setup has been described in detail elsewhere.[33] Very briefly, styrene (Fisher Scientific, 99.5%) was placed in a small vessel external to the photoelectron imaging spectrometer. Helium (0.4 bars) was passed through this liquid sample reservoir and then introduced into the spectrometer source chamber by continuous expansion through a small pinhole (∅ = 150 μm). After passing into the main interaction chamber via a skimmer, the molecular beam was intersected by co-propagating UV pump and probe pulses derived from the fundamental 800 nm output of a 1 kHz Ti:Sapphire laser system with a 55 fs full-width at half maximum (FWHM) Gaussian pulse width. The probe beam (267 nm, 1.0 µJ/pulse) was provided by the third harmonic of this output, using thin BBO crystals as the non-linear medium. This beamline also incorporated a computer-controlled linear translation stage for automated control of the temporal pump-probe delay. In this proof-of-principle setup, the pump beam (242-258 nm central wavelength) was generated inside a 17 cm long fused silica anti-resonant guiding, single-ring, HC-PCF with a 26 µm core (see Fig. 1 for structure) enclosed in a pressurized gas-cell. The wall thickness of the single-ring was 205 nm, providing a visible to infrared guidance band, along with UV transmission across 240 to 480 nm.[34] Other UV regions can be accessed with modified fibre designs. The coupled input 800 nm energy was 2.2 µJ/pulse and the fibre was filled with 4.5-6.0 bars of argon to tune RDW emission over the required wavelength range. Output energy in the UV was up to 50 nJ/pulse. Given the extremely short temporal duration and associated large (approx. 12 nm FWHM) spectral bandwidth, pulse propagation in air was restricted to just a few centimetres before passing via a thin (0.5 mm) $CaF_2$ window into a differentially pumped vacuum box coupled directly to the spectrometer. Once under vacuum, pump beam steering and combination with the probe were then undertaken.

Pump-probe ionisation took place between the electrodes of an electrostatic lens set-up optimised for velocity-map imaging (VMI).[35] The unfocused light pulses initially passed straight through the VMI set-up and were then incident on a curved UV enhanced aluminium mirror ($R$ = 10 cm). Tightly focussed beams then excited and ionized the styrene sample upon passing back through the VMI electrodes. A 40 mm MCP/P47 phosphor screen detector was used in conjunction with a CCD camera (640 × 480 pixels) to image the resulting photoelectrons. A pump-probe cross correlation of 160 ± 20 fs was obtained directly from non-resonant (1 + 1′) ionization of 1,3 butadiene and energy calibration data was obtained from three-photon, non-resonant ionisation of xenon using the 267 nm probe. Despite the large bandwidth of the pump, the spectrally resolved cross correlation measurement exhibited no significant chirp (upper panel of Fig. 3) indicating that dispersion effects were adequately controlled and the temporal experimental resolution was limited by the longer 267 nm pulse. The power and central wavelength output from the fibre remained stable for in excess of 100 hours of data acquisition, without the need for any laser pointing stabilization system.

A global fitting routine was used to model the time-dependent dynamics of the excited styrene molecules. Angle-integrated photoelectron spectra $S(E, \Delta t)$ were modelled by $n$ exponentially decaying functions $P_i(\Delta t)$, each convolved with the experimental cross-correlation $g(\Delta t)$

$$S(E, \Delta t) = \sum_{i=1}^{n} A_i(E) \cdot P_i(\Delta t) \otimes g(\Delta t). \quad (1)$$

The global fit returns the $1/e$ decay lifetime $\tau_i$ and the energy-dependent amplitudes $A_i(E)$ for each $P_i(\Delta t)$, providing a decay associated spectrum (DAS) attributable to a dynamical process operating on a specific timescale. This is a parallel model (i.e. all fit functions originate from $\Delta t = 0$), and any negative amplitude present in the DAS indicates a sequential dynamical process. Additional information and illustrative examples relating to this approach may be found elsewhere.[36, 37] The angular information in our VMI images was also analyzed using methodology described previously.[38, 39] Photoelectron angular distributions were found to be largely isotropic and displayed no significant temporal evolution. This aspect of our data will therefore not be considered further here.[40]

A time resolved photoelectron spectrum of styrene obtained using a 247 nm pump is presented in the lower panel of Fig. 3. As is evident from Fig. 2, here it is expected that the $S_2(\pi\pi^*)$ state will be predominantly excited.[41, 42] Ultrafast internal conversion will then lead to population of the energetically lower-lying $S_1(\pi\pi^*)$ state. The $S_2(\pi\pi^*)$ dynamics are reflected in the very short-lived spectral feature seen in the 0.6-1.3 eV photoelectron kinetic energy region. At very low (<0.6 eV) photoelectron kinetic energies, the spectrum is dominated by a very long-lived band that is attributed to ionization of the $S_1(\pi\pi^*)$ state. Greater insight is provided by fitting the experimental data using Eq. 1. To generate a satisfactory model, three exponentially decaying functions were required and we label these using their respective time constants $\tau_{1-3}$. The corresponding DAS are shown in Fig. 4. The short-lived (40 ± 10 fs) $\tau_1$ DAS exhibits positive amplitude above 0.6 eV and negative amplitude at lower photoelectron kinetic energies. This is highly characteristic of non-adiabatic population transfer from the $S_2(\pi\pi^*)$ state to the $S_1(\pi\pi^*)$ state.[36, 37] In the region below 0.6 eV, both $\tau_2$ (610 ± 60 fs) and $\tau_3$ (48 ± 5 ps) display positive amplitude and are attributed to ionization from $S_1(\pi\pi^*)$. Qualitatively similar DAS were obtained at all pump wavelengths, although at 258 nm $\tau_1$ was extremely weak as here the extent of direct $S_1(\pi\pi^*)$ excitation becomes much greater (see Fig. 2). A summary of the $\tau_{1-3}$ lifetimes obtained at all pump wavelengths is presented in Table I.

A previous TRPEI study by Nunn *et al.* has suggested that the $S_1(\pi\pi^*)$ state lifetime of styrene shows a marked dependence on whether it is prepared directly (i.e. optically) at 254 nm or via internal conversion following 240 nm excitation.[32] Somewhat surprisingly, the former scenario (with 0.6 eV of internal vibrational energy) was reported to give rise to a much shorter lifetime (4 ps) than the latter (19 ps, with 0.9 eV $S_1(\pi\pi^*)$ vibrational excitation). This was attributed to the very different molecular geometries required to access conical intersections connecting the $S_1(\pi\pi^*)$ state to $S_2(\pi\pi^*)$ and the $S_0$ ground state (into which the $S_1(\pi\pi^*)$ population ultimately decays). This is, however, not fully reconciled with an earlier study of Stolow and co-workers, who still observed ultrafast (50 fs) internal conversion between $S_2(\pi\pi^*)$ and $S_1(\pi\pi^*)$ following 254.3 nm excitation, in addition to a long $S_1(\pi\pi^*)$ lifetime of 84 ps.[31] Given the very large bandwidth of our 247 nm pump pulse, it is reasonable to assume that our data contains a non-negligible contribution from direct $S_1(\pi\pi^*)$ excitation as well as population of this state via non-adiabatic transitions.[41] This may then provide an explanation for the two distinct photoionization signals described by $\tau_2$ and $\tau_3$. In order to investigate this further, Fig. 5 plots the ratio of the energy-integrated $\tau_2/\tau_3$ DAS amplitudes at central pump wavelengths spanning the 258-242 nm region. This ratio would be expected to increase significantly as the pump wavelength gets longer if $\tau_2$ and $\tau_3$ originate from the decay of independent $S_1(\pi\pi^*)$ populations that are accessed directly and via internal conversion, respectively. As shown in Fig. 5, however, the $\tau_2/\tau_3$ ratio appears largely invariant with pump energy and so an alternative explanation for $\tau_2$ is required. In all cases we are populating $S_1(\pi\pi^*)$ well beyond the origin, which sits at 287.7 nm.[43, 44] Excitation at 258 nm is almost 0.5 eV (~4000 cm$^{-1}$) above this point, which comfortably exceeds the threshold for intramolecular vibrational energy redistribution (IVR) reported in the $S_1$ excited states of several small aromatic systems – including styrene itself.[45-47] One possibility for the $\tau_2$ DAS data is therefore sub-picosecond IVR within the $S_1(\pi\pi^*)$ state, leading to a modified Franck-Condon overlap with the cation and an associated decrease in ionization signal. An alternative picture yielding a similar outcome could also be the rapid dephasing of an initially localized vibrational wavepacket on the $S_1(\pi\pi^*)$ potential energy surface. Within the 3 ps linear time-step window of our experiment, however, no evidence of coherent oscillations (that might indicate a revival of such a wavepacket) were observed in any region of the various photoelectron spectra that were recorded.

Our experimental findings represent the first successful demonstration of rare-gas filled HC-PCFs as ultraviolet light sources for femtosecond pump-probe spectroscopy. The high efficiency and extremely broad tuneability this approach offers across the UV and VUV spectral regions has enormous future potential to enhance a wide range of ultrafast applications. In particular, accessing the VUV region spanning 195-165 nm has previously been extremely challenging using compact, bench-top sources. Furthermore, using two HC-PCF sources in tandem offers the possibility of exploiting the greatly improved (i.e. sub-20 fs) time resolution that the soliton-effect compression phenomenon inherently offers. We intend to begin exploiting these opportunities in the near future. We also anticipate that our initial validation of the HC-PCF approach will generate important stimulus in this area amongst the wider community, motivating uptake of a new technology that has genuine potential to transform ultrafast spectroscopy over the next decade.

**ACKNOWLEDGEMENTS**

This work was supported by EPSRC Grants EP/R030448/1 and EP/P001459/1, and by the European Research Council (ERC) under the European Union's Horizon 2020 research and innovation program: Starting Grant agreement HISOL, No. 679649.

**Table I**

| Central pump wavelength (nm) | $\tau_1$ (fs) | $\tau_2$ (fs) | $\tau_3$ (ps) |
|---|---|---|---|
| 242 | 60 ± 10 | 840 ± 170 | 45 ± 3 |
| 247 | 40 ± 10 | 610 ± 30 | 48 ± 4 |
| 253 | 70 ± 30 | 880 ± 150 | 59 ± 5 |
| 258 | <50 fs | 700 ± 280 | 94 ± 10 |

**Table I:** Summary of the time constants obtained for styrene at four different central pump wavelengths using the fitting procedure described by Eq.1.

**FIGURE CAPTIONS**

**Figure 1:** (a). Experimental set-up for ultrafast nonlinear gas experiments. A short (17 cm) length of HC-PCF was connected to two pressure cells, each with a built-in window. A scanning electron micrograph showing the cross-section of the fibre used for the present measurements is also included. (b). Numerical simulation illustrating the key features of deep-ultraviolet generation for 1 µJ, 30 fs FWHM pump pulses (centre wavelength 800 nm) in a HC-PCF with a 26 µm core diameter filled with 5 bar Ar. The minimum FWHM duration (1.4 fs) occurs at the point along the fibre where ultraviolet light is generated (~10 cm), where the intensity is a maximum. The colour scale is logarithmic. Adapted from Travers and co-workers.[23] (c). Demonstration of the broad tuning range possible by varying the gas and associated pressure. Adapted from Bang and co-workers.[27]

**Figure 2:** Room temperature styrene absorption spectrum recorded using a commercial bench-top UV spectrometer. The $S_1(\pi\pi^*)$ and $S_2(\pi\pi^*)$ bands are indicated and Gaussian functions illustrating the 242 nm and 258 nm pump central wavelength and associated bandwidth (~12 nm) are overlaid.

**Figure 3:** Time-dependent photoelectron spectra obtained using a 247 nm HC-PCF pump/267 nm probe. Top: Non-resonant (1+1′) ionization of 1,3 butadiene with pump-probe delays sampled in 30 fs increments. Bottom: Resonant (1+1′) ionization of styrene. Pump-probe delays between -450 fs and +450 fs were sampled in 30 fs increments, +450 fs to 3060 fs in 90 fs increments and a further 12 exponentially increasing steps then taken out to +200 ps. At each repeatedly sampled delay position, pump alone and probe alone images were also recorded for background subtraction. No probe-pump signals evolving towards negative time delays were observed due to the extremely weak intensity of the pump relative to the probe. For clear display of the dynamics, the time axis is linear to +3060 fs and then logarithmic out to +200 ps. The data are partitioned into 0.05 eV energy bins.

**Figure 4:** Decay associated spectra (DAS) obtained from a global multi-exponential fit to the styrene data presented in Fig. 3 using Eq. 1. Quoted uncertainties are 1σ values. Vertical dashed line denotes the predicted maximum photoelectron kinetic energy cut-off, based on the central pump (247 nm) and probe (267 nm) wavelengths and the adiabatic ionization potential (8.46 eV).[43, 48]

**Figure 5:** Ratio of the integrated $\tau_2$ and $\tau_3$ DAS amplitudes as a function of the central pump laser wavelength. Error bars denote 1σ uncertainties.

**Figure 1**

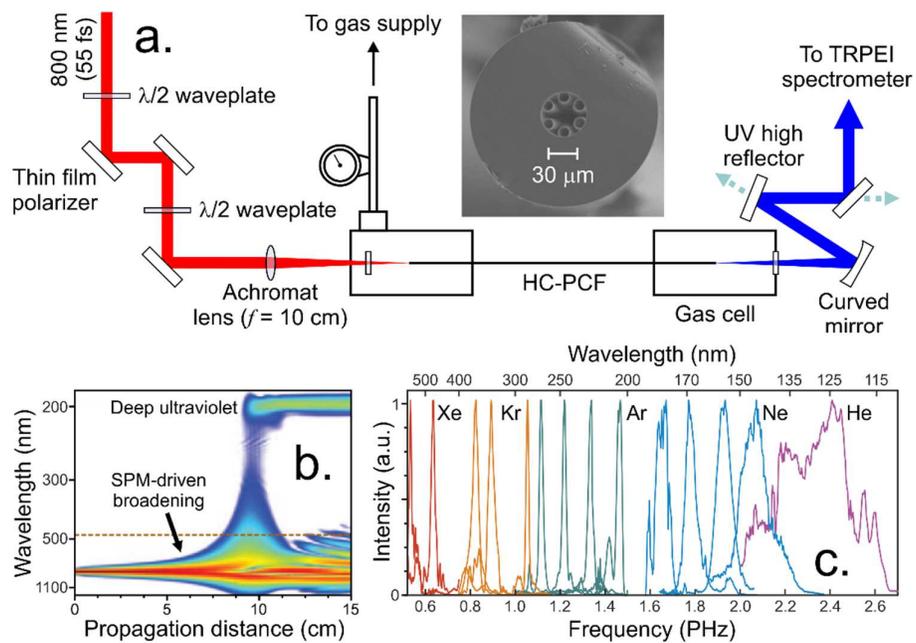

**Figure 2**

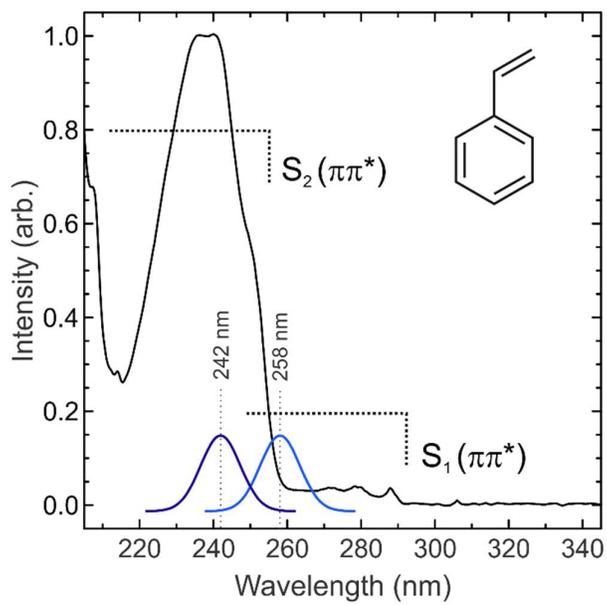

**Figure 3**

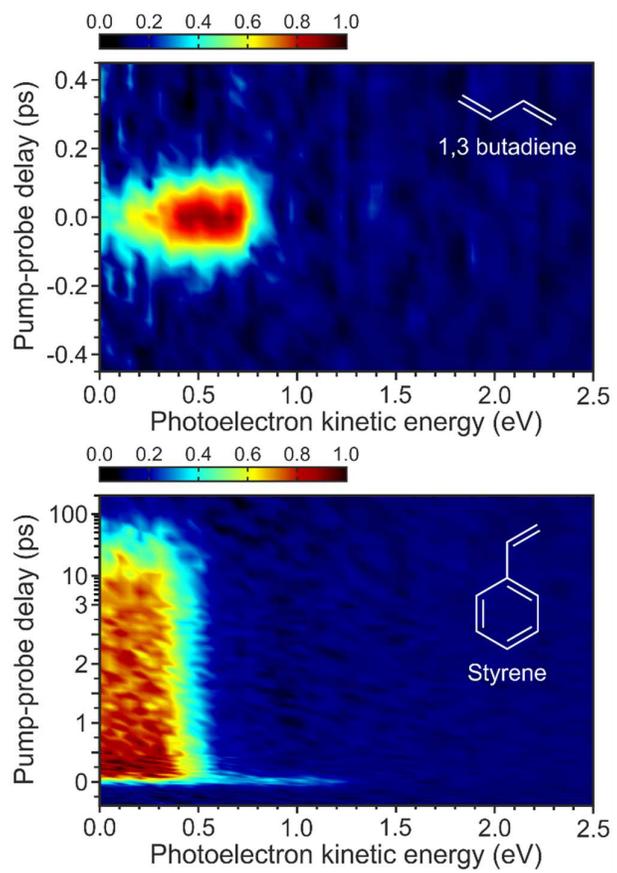

**Figure 4**

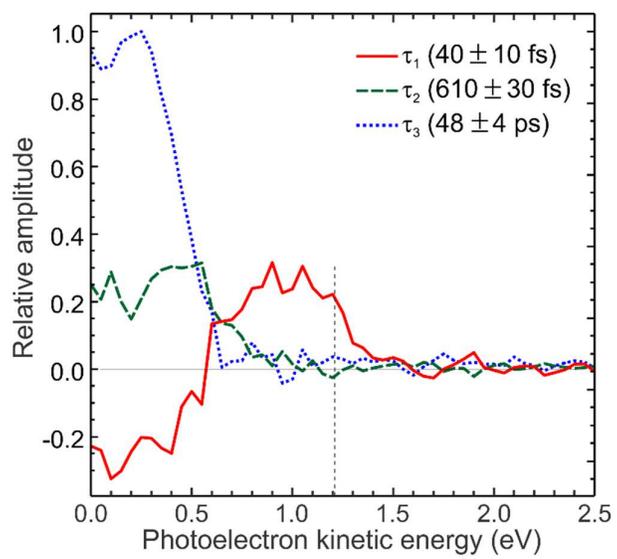

**Figure 5**

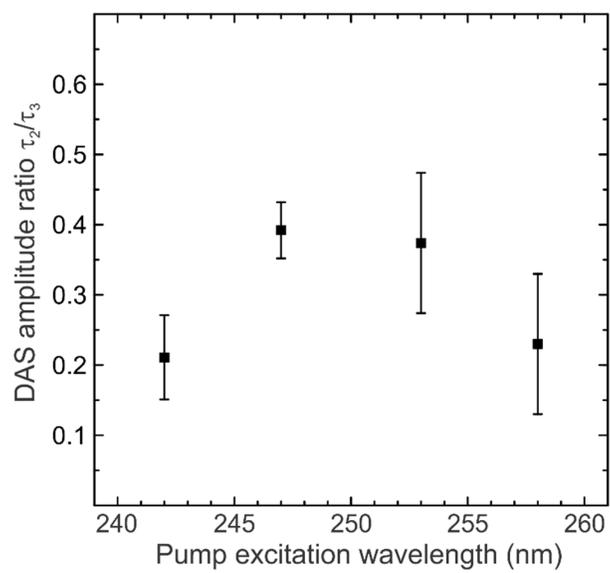